\documentclass[12pt]{article}
\usepackage{graphicx}
\usepackage{cite}
\usepackage{color}
\newcommand{\bm}[1]{\mbox{\boldmath $#1$}}
\newcommand{\abs}[1]{\left| #1\right|}
\newcommand{\fnd}[2]{\frac{\textstyle #1}{\textstyle #2}}
\newcommand{\fndrs}[4]{\fnd{\raisebox{#1}{$#2$}}{\raisebox{#3}{$#4$}}}
\newcommand{\dissum}[2]{\displaystyle \sum_{#1}^{#2}}
\newcommand{\braket}[3]{\mbox{$\left\langle #1\left|
#2\right| #3\right\rangle$}}
\newcommand{\Imag}[1]{\Im {\it m}(#1 )}
\newcommand{\xrm}[1]{{\textstyle \mbox{\rm #1}}}
\begin{document}
\title{Meson-meson interactions and Regge propagators}
\author{
Eef van Beveren$^{1}$ and George Rupp$^{2}$\\ [10pt]
{\small\it $^{1}$Centro de F\'{\i}sica Computacional,
Departamento de F\'{\i}sica,}\\
{\small\it Universidade de Coimbra, P-3004-516 Coimbra, Portugal}\\
{\small\it http://cft.fis.uc.pt/eef}\\ [10pt]
{\small\it $^{2}$Centro de F\'{\i}sica das Interac\c{c}\~{o}es Fundamentais,
Instituto Superior T\'{e}cnico,}\\
{\small\it Universidade T\'{e}cnica de Lisboa, Edif\'{\i}cio Ci\^{e}ncia,
P-1049-001 Lisboa, Portugal}\\
{\small\it george@ist.utl.pt}\\ [10pt]
{\small PACS number(s): 11.80.Gw, 11.55.Ds, 13.75.Lb, 12.39.Pn, 12.40.Nn}
}
\maketitle
\begin{abstract}
By a reformulation of the loop expansion
in the Resonance-Spectrum-Expansion amplitude for meson-meson scattering,
in terms of $s$-channel exchange of families of propagator modes,
we obtain a formalism which allows for a wider range of applications.
The connection with the unitarized amplitudes employed
in some chiral theories is discussed.
We also define an alternative for the Regge spectra and indicate
how this may be observed in experiment.
\end{abstract}

\section{Introduction}

In Ref.~\cite{PRD21p772}
a nonrelativistic Schr\"{o}dinger model
was proposed for the scattering amplitude
of non-exotic multi-channel meson-meson scattering,
which allows an exact solution in the form
of an analytic expression for the scattering amplitude.
Bound states and resonances are obtained
through the coupling of the two-meson system
to a harmonic oscillator (HO), the oscillator frequency being
independent of flavor.
By fine-tuning the intensity of the coupling,
one can transform the oscillator spectrum into
the spectrum of mesons for all possible flavor combinations.
The model's results, in particular concerning the resonance and
bound-state pole structure in the scalar-meson sector
\cite{ZPC30p615,PRL91p012003}, are well known.
Here, it is our aim to show that this model
corresponds to $s$-channel exchange of families of
propagator modes, similarly to the exchange
of a family of leading and daughter Regge trajectories.

In the naivest quark-model picture,
quarks and antiquarks are assumed to be confined
to a small region in space by strong forces.
The bulk of the interactions of the quarks with the glue
is contained in effective, or constituent, quark masses,
whereas the remaining dynamics is described
by a confining potential of sorts
\cite{PRL38p1309,KFKI1982d99}.
The quantum numbers of the effective $q\bar{q}$ system
are given by the total $q\bar{q}$ spin $S$,
the relative $q\bar{q}$ orbital angular momentum $L$,
and the $q\bar{q}$ radial excitation $N$.
For the lowest radial states of different flavor combinations,
i.e., having $N=0$, the confinement-model parameters can be
adjusted to experiment so as to obtain reasonable results
\cite{PRD48p4408}.  However, for higher radial excitations,
the results are poor \cite{PRD77p034501}.
Of course, the reason is the absence of meson loops, in combination
with a wrong fine-tuning of the model parameters
of the $N=0$ states to the lowest states in the experimental
spectrum \cite{PRD21p772}.

More elaborate quark models do include meson loops
\cite{PRL36p500,PRD21p772,PRL49p624,PRD42p1635},
and predict physical resonances
instead of fictitious real meson masses only.
On the other hand, in an effective, non-microscopic picture, Oller and Oset
\cite{NPA620p438} dynamically generated the low-lying scalar mesons
$f_0$(600) (alias $\sigma$ meson), $f_0$(980), and $a_0$(980) (the
$K_0^*$(800) or $\kappa$ meson was not studied),
by means of coupled-channel Lippmann-Schwinger equations
with meson-meson potentials resulting from the lowest-order chiral Lagrangian.
Their procedure boils down to using an effective four-meson vertex,
summing up the bubbles from the meson loops,
and unitarizing the resulting scattering amplitude.
Hence, they nicely showed that it only takes a four-point interaction
to generate scattering poles associated with light scalar mesons.
Earlier, T\"{o}rnqvist had made a similar suggestion
\cite{PRL49p624}, but went one step further
by also assuming {\it pre-existing} \/mesons,
which correspond to an {\it input spectrum for bare mesons}.
Unfortunately, his proposal ignored the issue of Adler zeros,
and so did not allow to conclude that in his model
actually two scalar-meson nonets can be generated.
Later, in Ref.~\cite{ZPC68p647} he did include Adler zeros,
but in the subsequent analysis, together with Roos \cite{PRL76p1575},
the $K^{\ast}_{0}(800)$ (alias $\kappa$ meson) was still not found,
probably due to the use of an unphysical,
negative Adler zero in the $I\!=\!1/2$ case \cite{AIPCP756p360}.
The latter pole of the isodoublet $S$-wave scattering amplitude
was then indeed generated by Oller and Oset, together with
Pel\'{a}ez \cite{PRD59p074001}, in an alternative coupled-channel
unitarization scheme.
Furthermore, in  Ref.~\cite{PRD60p074023} Oller and Oset
reported the observation of a second nonet, associated with
{\it pre-existing} \/meson states, in an $N/D$ approach to unitarization.
This issue was thoroughly studied
by Boglione \& Pennington in Ref.~\cite{PRD65p114010},
who came to the conclusion that it is indeed possible
to generate two states starting from one bare {\it seed} \/(or
{\it pre-existing} \/state) only,
and that it even may be plausible
to dynamically generate many states
with the same quantum numbers but different masses.
That is exactly what had been proposed almost two decades earlier
in Ref.~\cite{ZPC30p615}.

In Ref.~\cite{PRD21p772} an infinity of {\it seeds}
\/was introduced, all related through one parameter, viz.\
the oscillator frequency $\omega$,
being the same for all flavors.
When, at low energies, the sum is reduced to an effective constant,
one obtains a four-meson vertex which dynamically generates
exactly one pole in the case of the scalar mesons.
Also, at slightly higher energies, the sum can be approximated
by its leading term and an effective constant
for the remaining sum \cite{EPJC22p493}.
Then one obtains, apart from the dynamically generated resonance,
a second pole associated with the one leading {\it seed}.
However, in general, there is no need to approximate the sum,
thus allowing to generate several dynamical resonance poles,
\cite{ZPC30p615,PRL91p012003,AIPCP1030p219}
and moreover an infinite number of normal resonances
associated with the infinity of {\it seeds}.
Furthermore, the model of Ref.~\cite{PRD21p772}
can be applied to different flavos, including charm and bottom,
with just one set of parameters.
The low-lying scalar-meson nonet pops up without even being anticipated
\cite{ZPC30p615},
using the parameters that were fine-tuned
for the vector and pseudoscalar spectra in Ref.~\cite{PRD27p1527}.
However, there is an important difference
with the technique of introducing by hand one or more {\it seeds},
namely that the coupling constants can be controlled.

The problem of couplings and propagator modes is not new.
In Ref.~\cite{PLB25p475},
Delbourgo, Rashid, Salam, and Strathdee
remarked that it is well-known
{\it ``that Regge trajectories arise from sums of
infinite sequences of Feynman diagrams in conventional field theory''},
and, moreover, that this
{\it ``poses the problem of suitable coupling constants}''.
The latter problem was solved in Ref.~\cite{PLB25p475},
where a strategy was developed
for the decomposition of a Regge trajectory
in its propagator modes.
Here, we shall follow a different
but comparable approach.

In Refs.~\cite{PRD21p772,PRD27p1527},
the couplings of the propagator modes were controlled
by Clebsch-Gordans and 9-j symbols,
assuming the $^{3\!}P_{0}$ mechanism
\cite{NPB10p521,PRD2p336,PRD8p2223,AP124p61}
for quark-pair creation.
But this procedure led to inconsistencies
in the light-meson sector.
However, not only the couplings, but
also the choice of the spectrum of propagator modes
is of importance.
In Refs.~\cite{PRD21p772,PRD27p1527},
such a choice was made involving one free parameter,
viz.\ the universal level splitting,
which appeared to be largely flavor independent,
as we shall further elaborate in Sect.~\ref{propjesmodes} below.
Based on the choice of the spectrum,
the couplings issue
was solved by the $^{3\!}P_{0}$ recoupling strategy
of Refs.~\cite{PRD25p2406,ZPC17p135,ZPC21p291},
leaving only one parameter free, which
represents the probability of quark-pair creation.
As a consequence, there are only two free parameters,
no matter if the whole sum of propagator modes
is approximated by just one point vertex,
or by the sum of a few propagator modes and one point vertex.

The Schr\"{o}dinger equation
for the meson-meson scattering model
was solved with the wave-function approach (WFA)
in Refs.~\cite{PRD21p772,PRD27p1527}.
Later, in Ref.~\cite{IJTPGTNO11p179},
the same dynamical equations were solved
by an iterative method,
which we refer to by {\it Resonance-Spectrum Expansion} \/(RSE).
The latter method allows more easily a comparison with
current theories for meson-meson interactions.
However, it should be kept in mind that both approaches, WFA and RSE,
lead from the same dynamical equations
to the same expression for the meson-meson scattering amplitude.

In Refs.~\cite{ARXIV08050552,ARXIV08054803},
besides the usual four-meson interaction,
a fictitious bare scalar $\kappa$ meson was introduced,
in a similar fashion as suggested
by T\"{o}rnqvist \cite{PRL49p624}
and Oller \& Oset \cite{PRD60p074023},
leading to a two-pole description of the $K\pi$ $S$-wave interaction
for energies up to about 1.6 GeV.
The behavior of these poles,
as a function of one overal coupling parameter,
is comparable to the pole movements of the two lowest lying
$\kappa$ (or $K_0^*$) poles described in Refs.~\cite{ZPC30p615,EPJC22p493}.
Namely, the lower pole, which is dynamically generated by
the $K\pi$-$K\pi$ vertex
and associated with the $K^{\ast}_{0}(800)$,
moves towards larger negative imaginary energies,
away from the real axis, according as the coupling is decreased,
whereas the higher pole,
stemming from the {\it seed},
approaches the real axis for decreasing coupling,
ending up at the mass of the fictitious meson for vanishing coupling.
This similarity in pole behavior inspires us to revisit
the meson-meson scattering model of Ref.~\cite{PRD21p772},
and relate it to the exchange of Regge propagators
\cite{PLB25p475}.

\section{The scattering amplitude}

We define the amputated amplitude
for non-exotic two-meson scattering by
\begin{equation}
t=V+V\Omega V+V\Omega V\Omega V+\dots
=V\,\left[ 1-\Omega V\right]^{-1}
\;\;\; ,
\label{Texpansion}
\end{equation}
where $V$ stands for the RSE propagator
and $\Omega$ for the two-meson loop function.
In Ref.~\cite{IJTPGTNO11p179}, it was shown that the one-channel
RSE expression for the $\ell$-th partial-wave two-meson scattering
amplitude follows from
\begin{equation}
V_{\ell}(p)=\fnd{\lambda^{2}}{r_{0}}\,
\sum_{N=0}^{\infty}
\,\fnd{\abs{g_{NL}}^{2}}{E(p)-E_{NL}}
\;\;\;\;\xrm{and}\;\;\;\;
\Omega_{\ell}(p)=-2i\mu pr_{0}^{2}
j_{\ell}\left( pr_{0}\right)
h^{(1)}_{\ell}\left( pr_{0}\right)
\;\;\; ,
\label{Vandomega}
\end{equation}
where $p$ is the center-of-mass (CM) linear momentum,
$E(p)$ the total invariant two-meson mass,
$j_{\ell}$ and $h^{(1)}_{\ell}$
the spherical Bessel function and Hankel function of the first kind,
respectively, $\mu$ the reduced two-meson mass,
and $r_{0}$ a parameter with dimension mass$^{-1}$,
which can be interpreted as the average string-breaking distance.
The coupling constants $\lambda$ and $g_{NL}$ are discussed in
Sect.~\ref{MMMVertices},
where in Eq.~(\ref{Omatrix}) the relation
between $\ell$ and $L=L(\ell)$ is expressed.

\begin{figure}[htbp]
\begin{center}
\begin{tabular}{c}
\includegraphics[height=80pt]{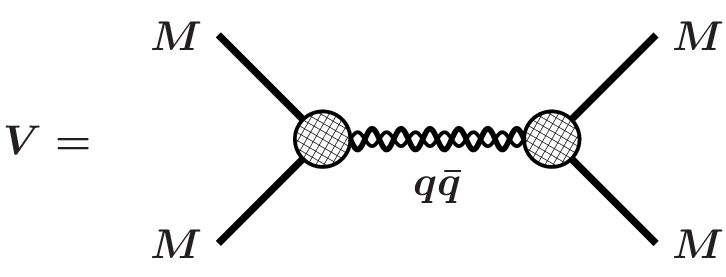}
\end{tabular}
\end{center}
\caption{Born diagram for {\it meson+meson$\,\to\,$meson+meson}.
The wiggly line represents the $s$-channel exchange of a system with
internal structure.}
\label{Kernel}
\end{figure}

The kernel $V$, which is graphically represented by Fig.~\ref{Kernel},
is the result of $s$-channel exchange
of a system with internal structure,
characterized by resonance modes
and their masses $E_{NL}$.
In the back of our mind we have, of course,
the picture of a resonating quark-antiquark system.
But this is not of much importance yet.
So we consider the exchange of propagators with structure,
rather than of single and pointlike massive objects,
in the very same spirit as a Regge propagator.

At the vertices,
each mode couples independently to the meson pairs.
The constants $g_{NL}$ indicate the intensities
of the couplings of the propagator modes to the meson pairs.
Now, since we consider an infinty of propagator modes,
we would end up with an infinite number of coupling parameters,
with which one could describe any two-meson system.
However, this freedom is strongly restricted by
the $^{3\!}P_{0}$ recoupling scheme \cite{ZPC17p135,ZPC21p291},
described in Sect.~\ref{MMMVertices}.
In fact, within the $^{3\!}P_{0}$ scheme,
all but one of the parameters $g_{NL}$ are fully determined.
The only remaining free parameter is $\lambda$, i.e.,
the overall three-meson-vertex coupling constant.

\begin{figure}[htbp]
\begin{center}
\begin{tabular}{c}
\includegraphics[height=80pt]{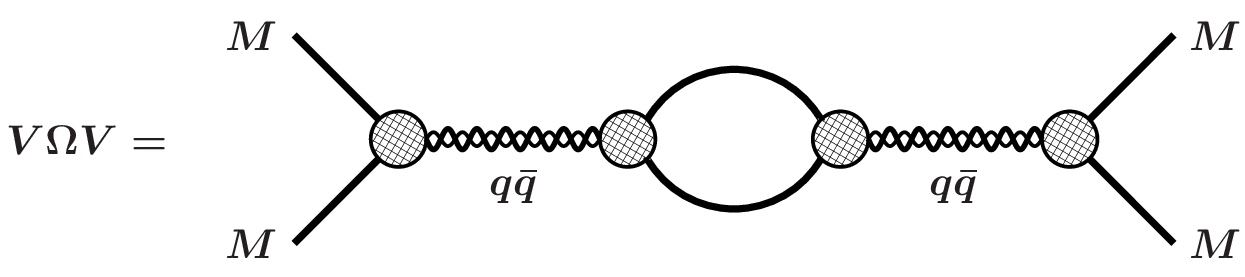}
\end{tabular}
\end{center}
\caption{Loop diagram for {\it meson+meson$\,\to\,$meson+meson}.
The wiggly line represents the $s$-channel exchange of a system with
internal structure. The mesons within the loop are stripped off their
internal dynamics and just carry the physical masses and quantum numbers.}
\label{loop}
\end{figure}

We thus assume that the RSE model can be formulated in terms of
$s$-channel exchanges of an infinite spectrum
of propagator modes.
The expression for the partial-wave loop function $\Omega_{\ell}(p)$
in Eq.~(\ref{Vandomega}) stems from the nonrelativistic
two-meson loops of expansion~(\ref{Texpansion}).
We have depicted $\Omega_{\ell}(p)$ in Fig.~\ref{loop}.
In the present formulation we are free to adopt different schemes
for the determination of loop functions.
This way, with not too much effort, the formalism
can also be extended to non-exotic baryon-meson
and baryon-antibaryon scattering and production processes.

Besides the coupling constants,
we also assume vertex functions
which regularize the loop integrations.
In the RSE, one employs a local spherical delta-shell
with radius $r_{0}$.
In the CM of the two-meson system,
its Fourier transform is given by a spherical Bessel function,
i.e.,
\begin{equation}
j_{\ell}\left( kr_{0}\right)
\;\;\; ,
\label{VertexFunctions}
\end{equation}
where $k$ stands for the CM loop momentum.
In the CM system, we furthermore adopt for the two-meson loops
the techniques developed by Logunov \& Tavkhelidze and
Blankenbecler \& Sugar \cite{PR142p1051,NC29p380},
and also follow the covariant prescription formulated by
Cooper and Jennings \cite{NPA500p553}.
This reduces the four-dimensional integration
to a three-dimensional one in the CM frame,
which, assuming spherical symmetry, takes the form
\begin{equation}
2\mu r_{0}^{2}
\int_0^{\infty} k^2dk\,\frac{j^2_{\ell}\left( kr_{0}\right)}
{p^2-k^2+i\epsilon}
\;\;\; ,
\label{loopdef}
\end{equation}
where $k$ and $p$ represent relativistic three-momentum moduli,
and $p^{2}$ is given by the on-shell relative two-meson momentum
in the CM frame.
This expression straightforwardly leads
to the loop function $\Omega_{\ell}(p)$ in Eq.~(\ref{Vandomega}).

One obtains a two-meson scattering amplitude $T$ satisfying unitarity,
from the amputated amplitude $t$ defined in Eq.~(\ref{Texpansion}),
by
\begin{equation}
T=X^{\dagger}tX=X^{\dagger}V\,\left[ 1-\Omega V\right]^{-1}X
\;,\;\;\;\xrm{where}\;\;\;\;
X^{\dagger}X=XX^{\dagger}=\Imag{\Omega}
\;\;\; .
\label{uniT}
\end{equation}
Consequently,
in the loop-expansion approximation of the scattering amplitude,
it is sufficient to determine the amputated propagator $V$
and the amputated loop function $\Omega$.
The latter, which is depicted in Fig.~\ref{loop},
consists here of two vertices and two free meson propagators.

The partial-wave amplitude $T_{\ell}(p)$
is related to the amputated partial-wave amplitude
$t_{\ell}(p)$ of Eq.~(\ref{Tpartial}) by
\begin{equation}
T_{\ell}=X^{\dagger}_{\ell}t_{\ell}X_{\ell}
=X^{\dagger}_{\ell}
V_{\ell}\,\left[ 1-\Omega_{\ell}V_{\ell}\right]^{-1}
X_{\ell}
\;,\;\;\;\xrm{where}\;\;\;\;
X^{\dagger}_{\ell}X_{\ell}=X_{\ell}X^{\dagger}_{\ell}=\Imag{\Omega_{\ell}}
\;\;\; ,
\label{uniTpartial}
\end{equation}
ensuring unitarity for each partial wave.

One ends up with an expression for the amplitude
of non-exotic two-meson scattering,
even in the case of many coupled channels,
which contains only three free parameters, viz.\
$\lambda$, $r_{0}$, and $\omega$,
besides the constituent quark masses.
The resulting amplitudes, for different flavors
and a variety of orbital quantum numbers,
have been confronted with experiment in numerous publications.
%\clearpage

\section{The propagator modes}
\label{propjesmodes}

Soon after T.~Regge noticed, in his famous work \cite{NC14p951},
that bound states and resonances of the scattering amplitude
are related to their poles in the complex orbital-angular-momentum plane,
Chew and Frautschi made the observation
that the squares of the masses of baryonic and mesonic resonances
come out on almost linear Regge trajectories \cite{PRL8p41}.
For mesons this has been explored in many models,
obtaining relations of the form \cite{HEPPH9912299}
\begin{equation}
E_{NL}^2=
C^{2}+2C\omega
\left( 2N+L+\frac{3}{2}\right)
\;\;\; ,
\label{quadraticReggeTraject}
\end{equation}
for meson masses $E_{NL}$.
Here, $N$ and $L$ represent
the radial and orbital-angular-momentum quantum numbers, respectively,
of the $q\bar{q}$ systems. Furthermore,
$C$ is a constant which depends
on the $q\bar{q}$ constituent flavor masses,
and $\omega$ is a universal frequency.
However, in Ref.~\cite{PRD21p772}
it was assumed that, at least for low energies,
the trajectories of propagator modes for $c\bar{c}$ and $b\bar{b}$
systems are linear, not quadratic, in their masses.
In Ref.~\cite{PRD27p1527}, the linear mass dependence was extended
to the light quarks $u$, $d$, and $s$, too,
while in Ref.~\cite{ZPC30p615}
the linear propagator modes were shown to explain the scalar resonances
in non-exotic $S$-wave meson-meson scattering as well as
the scattering data for energies up to about 2 GeV
in the $K\pi +K\eta +K\eta '$ complex
\cite{AIPCP814p143,AIPCP1030p219}.

For low excitations, the linear mass relation is obvious
from Eq.~(\ref{quadraticReggeTraject}) if
\begin{equation}
C\gg\omega
\;\;\; ,
\label{Cggomega}
\end{equation}
in which case one finds
\begin{equation}
E_{NL}\approx
C+\omega\left( 2N+L+\frac{3}{2}\right)
\;\;\; ,
\label{linearReggeTraject}
\end{equation}
but less obvious when $C$ and $\omega$
are of the same order of magnitude.
Now, for $\omega$ one finds in Ref.~\cite{PRD27p1527}
the value 0.19 GeV, whereas for $b\bar{b}$ and $c\bar{c}$
the values 9.45 GeV and 3.12 GeV, respectively, are quoted for $C$,
which satisfies well condition~(\ref{Cggomega}).
For the light quarks, $C\sim 1$ GeV,
which still is not in conflict with condition~(\ref{Cggomega})
for low energies.
Consequently, from the good results of the model
in Refs.~\cite{PRD21p772,PRD27p1527,ZPC30p615,AIPCP814p143,AIPCP1030p219},
one cannot exclude that, at high energies, the propagator trajectories
will be quadratic in mass.  However, since also the results for
the $K\pi +K\eta +K\eta '$ complex support linear relations,
we will stick here to relation~(\ref{linearReggeTraject})
for the model's Regge trajectories.

In previous work on the RSE, we referred to the modes of
the exchange propagator~(\ref{Vandomega})
as the confinement bound states.
This picture has not been completely abandoned here.
On the contrary, it will play an important role in determining
the parameters that describe the propagator modes.
Mode masses depend in the first place on the quark flavors
flowing in the propagator.
That information is contained in the constant $C$
of Eq.~(\ref{linearReggeTraject}).
However, we will argue in the following that the mode level splittings
are largely flavor independent.

The mesonic resonances extracted from experiment are organized by
flavor content, $J^{PC}I^{G}$ quantum numbers, mass and width.
Based on the $b\bar{b}$ and $c\bar{c}$ spectra,
it was concluded in Ref.~\cite{PRD21p772} that, in principle,
there must exist an infinity of such states, though most of the excited
states are difficult to observe because of the many open two-mesons
channels to which they couple.
albeit at higher masses obscured from observation
because of the many two-meson systems which couple to $q\bar{q}$.
Accordingly, we expect an infinite number of scattering poles
in meson-meson scattering, here represented by
\begin{equation}
E\; =\; P_{0}\, ,\;\; P_{1}\, ,\;\; P_{2}\, ,\;\;\dots
\;\;\; .
\label{polepositions}
\end{equation}
Unitarity then requires that in the one-channel restriction,
assuming the poles (\ref{polepositions})
to be simple,
the elastic scattering matrix $S$ be given by\footnote{
Note that we do not consider here
a possible overall phase factor
representing a background.} \cite{PRL49p624}
\begin{equation}
S(E)
\; =\;
\frac{\textstyle\raisebox{5pt}{
$\left( E-P_{0}^{\ast}\right)\left( E-P_{1}^{\ast}\right)
\left( E-P_{2}^{\ast}\right)\dots$}}
{\textstyle\raisebox{-5pt}{
$\left( E-P_{0}\right)\left( E-P_{1}\right)\left( E-P_{2}\right)\dots$}}
\;\;\; .
\label{unitairS}
\end{equation}

If we suppose that the resonances (\ref{polepositions})
stem from the spectrum of modes of propagator (\ref{Vandomega}),
given by the real quantities
\begin{equation}
E\; =\; E_{0}\, ,\;\; E_{1}\, ,\;\; E_{2}\, ,\;\;\dots
\;\;\; ,
\label{confinementpositions}
\end{equation}
then we may represent the differences $\left( P_{n}-E_{n}\right)$,
for $n=0$, 1, 2, $\dots$, by the complex mass shifts $\Delta E_{n}$.
Thus, we obtain for the unitary $S$-matrix the expression
\begin{equation}
S(E)
\; =\;
\frac{\textstyle\raisebox{5pt}
{$
\left( E-E_{0}-{\Delta E_{0}}^{\ast}\right)
\left( E-E_{1}-{\Delta E_{1}}^{\ast}\right)
\left( E-E_{2}-{\Delta E_{2}}^{\ast}\right)
\dots$}}
{\textstyle\raisebox{-5pt}
{$
\left( E-E_{0}-\Delta E_{0}\right)
\left( E-E_{1}-\Delta E_{1}\right)
\left( E-E_{2}-\Delta E_{2}\right)
\dots$}}
\;\;\; .
\label{unitairSdeltaE}
\end{equation}

So we assume here that resonances occur in scattering because
the two-meson system couples
to certain modes of the propagator (\ref{Vandomega}),
usually of the $q\bar{q}$ type,
viz.\ in non-exotic meson-meson scattering.
Let the strength of the coupling be given by $\lambda$.
For vanishing $\lambda$, we presume that the widths and real shifts
of the resonances also vanish. Consequently, the scattering poles
end up at the positions of the mode
spectrum~(\ref{confinementpositions}), and so
\begin{equation}
\Delta E_{n}\;\begin{array}{c} \\ \longrightarrow\\
\lambda\!\!\downarrow\! 0\end{array}\; 0
\;\;\;\;{\textstyle \mbox{\rm for}}\;\;\;\;
n\; =\; 0,\;\; 1,\;\; 2,\;\;\dots
\;\;\; .
\label{smalllambda}
\end{equation}
As a result, the scattering matrix tends to unity,
as expected in case there is no interaction.
The scattering amplitude (\ref{Texpansion})
exactly satisfies these requirements
for the propagator and loop function~(\ref{Vandomega}).
As a consequence, it seems that
one may only deduce an approximate mode spectrum from experiment.
Its precise masses $E_{NL}$ can then be found
by comparison to scattering and production data,
once the full scattering amplitude has been composed.
However, in Sect.~\ref{production} we shall see
that in production processes the Regge spectrum
may become visible.

In order to set out with the task to find a reasonable ansatz
for the mode spectrum of our propagator,
let us assume that the spectrum of mesonic quark-antiquark systems
can be described by flavor-independent HO confinement.
Then, for each pair of flavors, an infinite set of mesons exists with all
possible spin, angular, and radial excitations.
But unfortunately, for most flavor pairs only a few angular and even fewer
radial recurrencies are known \cite{JPG33p1}.
If we do not distinguish up and down,
but just refer to non-strange ($n$) quarks,
then we have at our disposal four different flavors:
$n$, $s$, $c$, and $b$.
These can be combined into ten different flavor pairs,
each of which may come in two different spin states: 0 or 1.
This gives rise to, in principle, twenty different meson spectra.
With some 150 known mesons, this means 7.5 angular plus radial
excitations on average, per flavor pair.
This is much less than e.g.\ the
known excitations of the positronium spectrum.
No wonder that it requires some imagination
to guess economic strategies for the description of mesons.

\begin{figure}[htbp]
\begin{center}
\begin{tabular}{c}
\includegraphics[height=160pt]{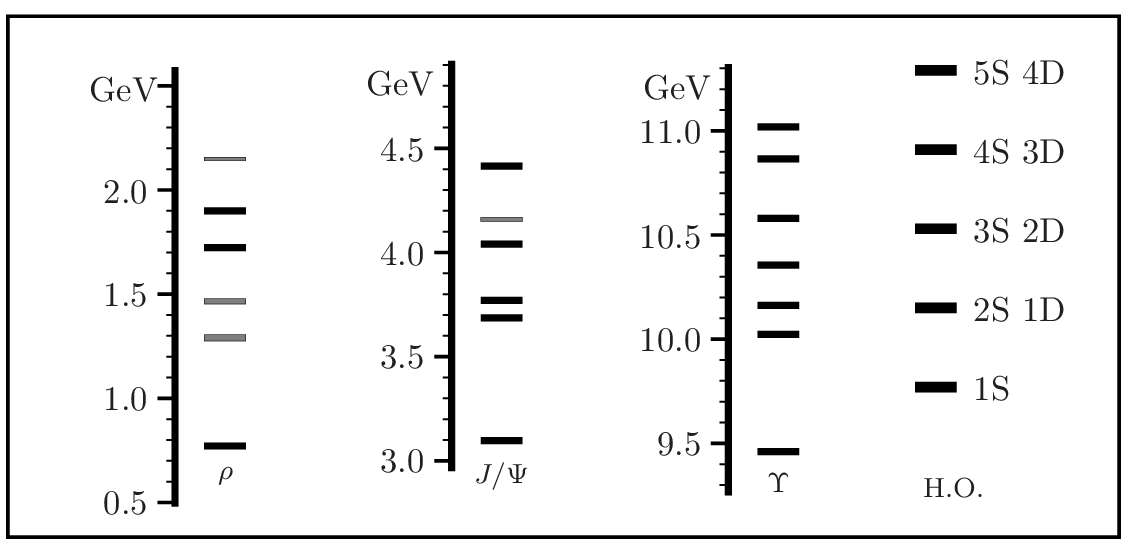}
\end{tabular}
\end{center}
\normalsize
\caption[]{Nonstrange, charmonium, and bottomonium $J^{PC}=1^{--}$ states
compared to the corresponding states from a harmonic-oscillator spectrum.
The level spacing for the oscillator equals 0.38 GeV.}
\label{vectors}
\end{figure}
As one may verify from the latest Review of Particle Physics
\cite{JPG33p1},
known vector states are more numerous than any other type of mesonic
resonances, since they are easier to produce.
Consequently, in order to structure
the mode masses, we begin with the vector mesons,
carrying quantum numbers $J^{PC}=1^{--}$.
In Fig.~(\ref{vectors}) we compare
the observed $n\bar{n}$, $c\bar{c}$, and $b\bar{b}$ vector states
with the HO possible states.
Most of the data are taken from Ref.~\cite{JPG33p1}.
The $\rho$(1250--1290) signal was originally reported in
Refs.~\cite{NPB76p375,NCA49p207,JINRP286682,NPPS21p105,SLACPUB5606},
and has very recently been confirmed in a coupled-channel
data analysis \cite{NPA807p145}.
The $\Upsilon (1D)$ has been observed in Ref.~\cite{HEPEX0207060}.

The charmonium vector states, shown in Fig.~(\ref{vectors}),
bear many similarities with the two-particle HO:
a ground state in a $c\bar{c}$ $S$-wave,
and higher $c\bar{c}$ radial excitations
that are {\it almost} \/degenerate
with the $c\bar{c}$ $D$-wave states.
Also, except for the ground state, the level spacings are roughly equal.
In Refs.~\cite{PRD21p772,PRD27p1527,HEPPH0201006},
the mechanism was discussed which turns
the HO spectrum into the charmonium spectrum,
also including the ground-state levels.

For the $\rho$ and $\Upsilon$ vector states,
also shown in Fig.~(\ref{vectors}), we see a very similar pattern:
the $q\bar{q}$ $S$-$D$ splittings are slightly larger,
while the $\rho$(770) ground state of the $\rho$ spectrum
and the $\Upsilon(1S)$ ground state of the $\Upsilon$ spectrum
also come out far below the corresponding oscillator ground states.
From Fig.~(\ref{vectors}) one may moreover conclude that,
as far as the level splittings are concerned,
there is not much reason
to separate the light-quark sector from the heavy quarks.
The mechanism which turns the oscillator states into
the $\rho$ and $\Upsilon$ resonances
is discussed in Ref.~\cite{PRD27p1527}.

What we learn from the above comparison is that
the level splittings are largely flavor independent and more or less
constant, whereby the ground-state level of the mode spectrum
is determined by the effective flavor masses.

\begin{table}[htbp]
\begin{center}
\begin{tabular}{||c||cc|c|cccc||}
\hline\hline & & & & & & & \\ [-7pt]
& \multicolumn{2}{c|}{$I=1$} & $I=\frac{1}{2}$ &
\multicolumn{4}{c||}{$I=0$} \\
& & & & & & & \\ [-7pt]
& \multicolumn{2}{c|}{$n\bar{n}$} & $n\bar{s}$ &
\multicolumn{2}{c}{$n\bar{n}$} &
\multicolumn{2}{c||}{$s\bar{s}$} \\
& & & & & & & \\ [-7pt]
\hline & & & & & & & \\ [-7pt]
$0^{+}$ &
$a_{0}(1450)$ & &
$K_{0}(1430)$ & $f_{0}(1370)$ & & $f_{0}(1500)$ & \\
& & &
$K_{0}(1980)$ & $f_{0}(1710)$ & & $f_{0}(2020)$ & \\
& & & & $f_{0}(2200)$ & & & \\
& & & & & & & \\ [-7pt]
\hline & & & & & & & \\ [-7pt]
$1^{+}$ &
$a_{1}(1260)$ & $b_{1}(1235)$ &
$K_{1}(1270)$ &
$f_{1}(1285)$ & $h_{1}(1170)$ & $f_{1}(1420)$ & $h_{1}(1380)$ \\
&
$a_{1}(1640)$ & &
$K_{1}(1400)$  &
$f_{1}(1510)$ & $h_{1}(1595)$ & & \\
& & &
$K_{1}(1650)$ & & & & \\
& & & & & & & \\ [-7pt]
\hline & & & & & & & \\ [-7pt]
$2^{+}$ &
$a_{2}(1320)$ & &
$K_{2}(1340)$ & $f_{2}(1270)$ & & $f_{2}(1430)$ & \\
&
$a_{2}(1700)$ & & &
$f_{2}(1525)$ & $f_{2}(1565)$ & $f_{2}(1640)$ & $f_{2}(1810)$ \\
& & &
$K_{2}(1980)$ &
$f_{2}(1910)$ & $f_{2}(1950)$ & $f_{2}(2010)$ & $f_{2}(2150)$ \\
& & &
& $f_{2}(2300)$ & $f_{2}(2340)$ & & \\ [3pt]
\hline\hline
\end{tabular}
\end{center}
\caption[]{\small The experimentally observed light positive-parity mesons.}
\label{pospar}
\end{table}
In Table~\ref{pospar}, we show the experimental spectrum of light
positive-parity mesons.
Only for the $f_2$ states, almost enough resonance data are available,
to allow for comparison with HO confinement.
We therefore assume that the states in the first two $f_2$ columns
contain mostly non-strange $q\bar{q}$ pairs, and in the next two
predominantly $s\bar{s}$ pairs.
Furthermore, the quark pair may come
in a relative $P$-wave
(1\raisebox{1.0ex}{\scriptsize st} and
3\raisebox{1.0ex}{\scriptsize rd} column) or
$F$-wave
(2\raisebox{1.0ex}{\scriptsize nd} and
4\raisebox{1.0ex}{\scriptsize th} column).
From the values for the central mass positions as given in
Ref.~\cite{JPG33p1}, we collect in Table~\ref{massdiff}
mass differences for a selected set \cite{EPJA31p468} of $f_{2}$ states.
\begin{table}[htbp]
\begin{center}
\begin{tabular}{||c|c||}
\hline\hline & \\ [-7pt]
states & mass difference \\
& \\ [-7pt]
\hline & \\ [-7pt]
$m\left( f_{2}(1910)\right) - m\left( f_{2}(1525)\right)$ &
0.39 $\pm$ 0.01 GeV \\ [5pt]
$m\left( f_{2}(2300)\right) - m\left( f_{2}(1910)\right)$ &
0.38 $\pm$ 0.03 GeV \\ [5pt]
$m\left( f_{2}(1950)\right) - m\left( f_{2}(1565)\right)$ &
0.40 $\pm$ 0.02 GeV \\ [5pt]
$m\left( f_{2}(2340)\right) - m\left( f_{2}(1950)\right)$ &
0.39 $\pm$ 0.04 GeV \\ [5pt]
$m\left( f_{2}(2010)\right) - m\left( f_{2}(1640)\right)$ &
0.38 $\pm$ 0.05 GeV \\ [5pt]
$m\left( f_{2}(2150)\right) - m\left( f_{2}(1810)\right)$ &
0.34 $\pm$ 0.02 GeV \\ [5pt]
\hline\hline
\end{tabular}
\end{center}
\caption[]{\small The experimentally \cite{JPG33p1}
observed mass differences for isoscalar
light positive-parity mesons with $J=2$.}
\label{massdiff}
\end{table}
For the spectra of Fig.~\ref{vectors}
we deduced a level spacing of 0.38 GeV,
which agrees well with the splittings in Table~\ref{massdiff}.
As a first approximation, it thus seems reasonable
to adopt for the masses of the propagator modes
the expression
\begin{equation}
M\left( f,\bar{f};N,L\right)
=m_{f}+m_{\bar{f}}+\omega
\left( 2N+L+\frac{3}{2}\right)
\;\;\; .
\label{HOspectrum}
\end{equation}
Here, $f$ and $\bar{f}$ represent the flavors
of the quark and the antiquark,
$m_{f}$ and $m_{\bar{f}}$ their respective masses,
and $\omega$ the oscillator frequency.

In the following, let us study some details of formula (\ref{HOspectrum})
The vector-meson states have unit total angular momentum, $J=1$, and
unit $q\bar{q}$ total spin, $S=1$.
Hence, since the parity of vector-meson states equals $P=-1$,
their orbital angular momentum can be $L=0$ ($S$-wave) or
$L=2$ ($D$-wave).
From formula~(\ref{HOspectrum}) we then understand that, for
HO confinement, the vector-meson states
with ($N$,$L=2$) are degenerate with the vector-meson states
with ($N+1$,$L=0$), as shown in Fig.~(\ref{vectors}).
For other flavor and spin excitations similar results emerge.
One obtains a very regular,
equally spaced spectrum of propagator modes,
with an oscillator frequency $\omega$, which comes out at about
0.19 GeV for the data.

Non-strange ($n\bar{n}$) and strange ($s\bar{s}$) configurations
double the number of isoscalar states into $SU(3)$-flavor singlets
and octets.
But one should be be aware that all states are mixed through the
meson loops.
Hence, like in Nature we will not find
pure angular, radial, or flavor excitations
for resonances of the scattering amplitude (\ref{Texpansion}).
Lattice calculations reveal that it may even be
very hard to disentangle the various configurations showing up
in $f_{0}$ systems \cite{HEPLAT0210012}.
Moreover, meson loops influence the precise resonance shapes.
Some come out broad, others narrower,
Also, the central resonance positions may shift substantially
(100--300 MeV \cite{PRD21p772,PRD27p1527,ZPC30p615})
with respect to the propagator mode spectrum.

Now, as discussed in the beginning of this section,
we may not exclude the possibility that
the trajectories for the propagator modes
are quadratic in mass.
Hence, we may thus very well assume that relation (\ref{HOspectrum})
is only an approximation, valid for low energies,
and refer to propagators (\ref{Kernel})
as Regge Green's functions \cite{PLB25p475}.
We leave the study of a more precise relation with string theory
\cite{NCA57p190} for future research.
%\clearpage

\section{Vertices}
\label{MMMVertices}

In this section we study how the propagator modes
couple to the meson pair.
Thereto, we characterize each meson by its quantum numbers.
This is in part guess work,
since we only have at our disposal the total spin $J$,
the parity $P$, and, for flavorless mesons, the $C$-parity.
But let us suppose here that we also have knowledge
of the orbital angular momentum $L$, the internal spin $S$,
and the radial excitation $n$
of the quark pair that constitutes the meson.
Hence, a two-meson system then consists of
the sets of quantum numbers
$\left( J_{1},L_{1},S_{1},n_{1}\right)$
and $\left( J_{2},L_{2},S_{2},n_{2}\right)$,
characterizing each meson,
and also the quantum numbers describing
the relative motion of the two mesons, viz.\
$J$, $\ell$, $s$, and $n$.
The propagator modes are similarly characterized by
a set of meson quantum numbers
$\left( J,L,S,N\right)$.
Hence, the complete coupling is given by the matrix element
of the transition operator $\cal O$
\begin{equation}
\braket
{\left( J_{1},L_{1},S_{1},n_{1}\right) ;
\left( J_{2},L_{2},S_{2},n_{2}\right) ;
J,\ell,s,n}
{\cal O}
{J,L,S,N}
\;\;\; .
\label{Omatrix}
\end{equation}

Given this form of the coupling constants,
it is advantageous to determine the scattering amplitude
from the partial-wave expansion.
In the two-meson CM system,
assuming spherical symmetry, we define
\begin{equation}
T\left({\vec{p}\;}\right)\; =\;
\sum_{\ell =0}^{\infty}(2\ell +1)\;
P_{\ell}\left(\hat{p}\cdot{\hat{p}\,}'\;\right)\;
T_{\ell}(p)
\;\;\; .
\label{Tpartial}
\end{equation}

An important ingredient for the vertices employed
in the WFA and the RSE
is the $^{3\!}P_{0}$ mechanism for quark-pair creation
of Micu \cite{NPB10p521}
and Carlitz \& Kislinger \cite{PRD2p336},
which has been worked out in more detail
by Le Yaouanc, Oliver, P\`{e}ne \& Raynal \cite{PRD8p2223},
Chaichian \& K\"{o}gerler \cite{AP124p61},
and later by Ribeiro \cite{PRD25p2406}.
A complete expression of the latter matrix elements
for all possible quantum numbers
and different effective quark masses is given
in Refs.~\cite{ZPC17p135,ZPC21p291}.
The resulting couplings have been employed for perturbative
calculus of particle widths and mass shifts
in Refs.~\cite{APA11p171,NPA683p425,PLB548p153,PRC77p055206}.
In the latter works, flavor mass
and the universal oscillator frequency
were combined to one parameter,
different for each flavor.

In Ref.~\cite{IJTPGTNO11p179} it was shown that the
{\it spectral representation} \/of the Green's function, viz.\
\begin{equation}
\sum_{N=0}^{\infty}
\fnd{\abs{{\cal F}_{NL}\left( r_{0}\right)}^{2}}
{E-E_{NL}}\; =\;\fnd{2\mu}{r_{0}^{2}}\;
\fnd{F_{L}\left( E,r_{0}\right)G_{L}\left( E,r_{0}\right)}
{W\left(F_{L}\left( E,r_{0}\right),
G_{L}\left( E,r_{0}\right)\right)}
\;\;\; ,
\label{Complitude}
\end{equation}
connects the WFA expression for the full propagator
(righthand side \cite{PRD21p772})
with the RSE iterative result (lefthand side \cite{IJTPGTNO11p179}).
The full propagator is obtained by constructing
an exact solution of a nonrelativistic stationary equation
for meson-meson scattering containing a confining part $H_{c}$.
The expansion in eigensolutions of $H_{c}$
is obtained by an iterative method.
The set of functions
$\left\{{\cal F}_{NL}\; ;\; N=0,1,2,\dots\right\}$
represents, for orbital angular momentum $L$,
a full set of radial eigensolutions, with eigenvalue $E_{NL}$,
of the confining part $H_{c}$ of the full Hamiltonian.
Furthermore, $F_{L}$ and $G_{L}$ represent two linearly
independent solutions of $H_{c}$ for any value of the energy $E$.

The mode distribution in the CM frame
of the two-meson system is contained in
$\abs{{\cal F}_{NL}\left( r_{0}\right)}^{2}$
of Eq.~(\ref{Complitude}).
In the RSE, the mode distribution is contained in
the matrix elements of Eq.~(\ref{Omatrix}).
Henceforth,
we put aside the WFA description of the propagator modes
and its coupling to meson pairs,
and concentrate on the iterative RSE description.
Hence, when we abbreviate the result of matrix
element~(\ref{Omatrix}) for two-meson channel $i$ by
$g_{i,\ell}(N,L)$, then we propose here to write
for the partial-wave Regge propagator
connecting the two-meson channels $i$ and $j$
the expression
\begin{equation}
\left[ V_{\ell}(s)\right]_{ij}=\fnd{\lambda^{2}}{r_{0}^{2}}\,
\sum_{L=0}^{\infty}
\sum_{N=0}^{\infty}
\,\fnd{g_{i,\ell}(N,L)g_{j,\ell}(N,L)}{E-E_{NL}}
\;\;\; ,
\label{Vmoregeneral}
\end{equation}
where $L=L(\ell )$ and
in which form the RSE formalism takes a shape similar to
the result of Ref.~\cite{PLB25p475}.
The sum in $L$ is usually very much restricted,
since most couplings $g_{i,\ell}(N,L)$ vanish,
as follows from the details \cite{ZPC17p135,ZPC21p291}
of Eq.~(\ref{Omatrix}).
%\clearpage

The vertices connecting the multi-mode propagator
to the two-meson systems,
are calculated in the CM system.
On a basis of HOs,
for each mode one assumes a spatial distribution
of two quarks and two antiquarks,
one colorless pair for the propagator mode,
and an equally colorless $^{3\!}P_{0}$ pair.
These spatial distribution functions are decomposed
on the basis of two-meson distributions \cite{EPJC11p717}.
The coefficients of such a decomposition are called
{\it recoupling constants} \cite{ZPC17p135,ZPC21p291}.
In Table~\ref{somecouplings}
we have collected some of the $^{3\!}P_{0}$ recoupling constants,
merely as a demonstration.
\begin{table}[htbp]
\begin{center}
\begin{tabular}{||c|c||}
\hline\hline & \\ [-10pt]
vertex &  recoupling coefficients\\
${_{n}J}^{PC}\to{_{r}J}^{PC}+{_{r}J}^{PC}$ in
($\ell$,$s$)&
$\{ g(n)\}^{2}\times 4^{n}$\\
\hline & \\ [-10pt]
${_{n}0}^{-+}\to{_{0}0}^{-+}+{_{0}0}^{++}\;\;$ (0,0) &
$\frac{1}{144}(2n-3)^{2}$\\ [3pt]
${_{n}0}^{-+}\to{_{0}1}^{--}+{_{0}1}^{+-}\;\;$ (0,0) &
$\frac{1}{144}(2n-3)^{2}$\\ [3pt]
${_{n}0}^{-+}\to{_{0}1}^{--}+{_{0}1}^{++}\;\;$ (0,0) &
$\frac{1}{72}(2n-3)^{2}$\\ [3pt]
${_{n}0}^{-+}\to{_{0}0}^{-+}+{_{0}1}^{--}\;\;$ (1,1) &
$\frac{1}{24}(2n+3)$\\ [3pt]
${_{n}0}^{-+}\to{_{0}1}^{--}+{_{0}1}^{--}\;\;$ (1,1) &
$\frac{1}{12}(2n+3)$\\ [3pt]
\hline & \\ [-10pt]
${_{n}0}^{++}\to{_{0}0}^{-+}+{_{0}0}^{-+}\;\;$ (0,0) &
$\frac{1}{24}(n+1)$\\ [3pt]
${_{n}0}^{++}\to{_{0}0}^{-+}+{_{1}0}^{-+}\;\;$ (0,0) &
$\frac{1}{288}(2n+3)(n-1)^{2}$\\ [3pt]
${_{n}0}^{++}\to{_{1}0}^{-+}+{_{1}0}^{-+}\;\;$ (0,0) &
$\frac{1}{6\times 24^{2}}\, n(2n+1)(2n+3)(n-3)^{2}$\\ [3pt]
${_{n}0}^{++}\to{_{0}0}^{-+}+{_{0}1}^{++}\;\;$ (1,1) &
$\frac{1}{12}$\\ [3pt]
${_{n}0}^{++}\to{_{0}1}^{--}+{_{0}1}^{--}\;\;$ (0,0) &
$\frac{1}{72}(n+1)$\\ [3pt]
${_{n}0}^{++}\to{_{0}1}^{--}+{_{0}1}^{--}\;\;$ (2,2) &
$\frac{1}{18}(2n+5)$\\ [3pt]
${_{n}0}^{++}\to{_{0}1}^{--}+{_{1}1}^{--}\;\;$ (0,0) &
$\frac{1}{4\times 6^{3}}(2n+3)(n-1)^{2}$\\ [3pt]
${_{n}0}^{++}\to{_{0}1}^{--}+{_{0}1}^{--}_{\ell =2}\;\;$ (0,0) &
$\frac{1}{5\times 6^{3}}(2n+3)(2n-5)^{2}$\\ [3pt]
${_{n}0}^{++}\to{_{0}1}^{--}+{_{0}1}^{+-}\;\;$ (1,1) &
$\frac{1}{12}$\\ [3pt]
${_{n}0}^{++}\to{_{0}0}^{++}+{_{0}0}^{++}\;\;$ (0,0) &
$\frac{1}{432}(2n+3)(n-3)^{2}$\\ [3pt]
${_{n}0}^{++}\to{_{0}1}^{++}+{_{0}1}^{++}\;\;$ (0,0) &
$\frac{1}{144}(2n+3)(n-2)^{2}$\\ [3pt]
${_{n}0}^{++}\to{_{0}1}^{+-}+{_{0}1}^{+-}\;\;$ (0,0) &
$\frac{1}{144}(2n+3)(n-1)^{2}$\\ [3pt]
\hline\hline
\end{tabular}
\end{center}
\caption[]{\small Vertex recoupling constants $g(n)$
for the radial excitations $n$ ($n=0$, $1$, $2$, $\dots$)
of pseudoscalar and scalar modes of the propagator
(\ref{Kernel}), for the case of equal effective quark masses
\cite{ZPC17p135,ZPC21p291}.
We have characterized the two-meson systems by the quantum numbers
${_{r}J}^{PC}$ ($r$ for radial excitation),
the relative two-meson angular momentum $\ell$,
and the total two-meson spin $s$.
In one case, we have also indicated
the $q$-$\bar{q}$ internal angular momentum of the meson.
For the other cases, the lowest possible quantum numbers are assumed.}
\label{somecouplings}
\end{table}

For $n=0$, the recoupling constants squared in one column add up to 1,
with the proviso that meson pairs with two different mesons
count twice, as actually we should have repeated
the corresponding lines in the table for the interchanged pair.
This result reflects the fact that we consider
a properly normalized distribution for each mode,
and an orthonormal set of two-meson distributions.

For $n>0$, the recoupling constants squared in one column
do not add up to 1.
The reason is that for $n>0$ more possible two-meson systems
couple to the propagator modes.
If those were included in the tables,
we would obtain unity for all $n$.
A full table would have infinite length.
Moreover, even when limited to $J<10$ and only for a few radial excitations,
one would easily end up with a table of hundreds of pages.
However, with a fast computer code in Fortran,
based on the expressions
given in Refs.~\cite{ZPC17p135,ZPC21p291},
and which moreover takes care
of the various possible isospin combinations
and also allows for unequal effective flavor masses,
the absence of such tables is no limitation.
%\clearpage

\section{Comparison to other models}

In their comment \cite{PRL77p2332}
on the work of T\"{o}rnqvist \& Roos \cite{PRL76p1575},
Isgur \& Speth pointed out that,
since exotic channels do not couple to the propagator of
Eq.~(\ref{Vandomega}),
the corresponding scattering amplitude vanishes,
which is not in agreement with experiment.
Furthermore, they argued
that in the J\"{u}lich model \cite{PRD52p2690}
$t$-channel processes lie at the origin
of a {\it broad} \/dynamically generated pole
in the $I=0$ $S$-wave pion-pion scattering amplitude,
and not $s$-channel propagator modes.
Finally, they remarked that also the $a_{0}(980)$ and $f_{0}(980)$
are largely due to $t$-channel forces.

Now, it seems indisputable that with zero couplings
their is no scattering. Nevertheless, the first observation of
Isgur \& Speth is incomplete.

Namely, when just arbitrary couplings and {\it seeds} \/are involved,
then there is no clear prescription how to handle exotic channels.
However, when propagator modes and couplings are related,
like in the choice of Ref.~\cite{PRD21p772}
and the corresponding recoupling constants of
Ref.~\cite{ZPC21p291},
then quark-interchange processes, and so scattering in exotic channels,
make part of the interactions that can be handled in the model.

In Ref.~\cite{ZPC21p291},
the quark-interchange diagrams were suppressed
in order to single out the effect of the propagator modes
in $s$-channel exchange via $^{3\!}P_{0}$ pair creation.
Consequently, although in principle there is no limitation in the model
inhibiting the study of exotic channels,
it is just neglected in the RSE approach to non-exotic meson-meson scattering.
A study also including the quark-interchange diagrams
has been carried out by Bicudo and Ribeiro
in Ref.~\cite{ZPC38p453}.

Moreover,
in Ref.~\cite{PRL78p1603}
Fariborz, Jora \& Schechter
observed that neglecting the $t$-channel exchange of a $\rho$ meson
in $I=0$ $S$-wave $\pi\pi$ scattering does not remove the $\sigma$ pole
from the scattering amplitude of their model,
and only shifts it in a modest way.
Also note that the dynamical ``$\sigma(400)$'' pole generated
by the J\"{u}lich group \cite{PRD52p2690} is considerably lighter and
broader than generally accepted.

In Ref.~\cite{PRD65p114010}, also Boglione \& Pennington remarked:
{\it ``$s$-channel dynamics is not all that controls the scattering''},
when refering to Ref.~\cite{PRL76p1575}.
However,
one may not conclude from spicing up with $t$-channel exchange,
a model which only accounts for one (or a few) propagator modes,
moreover in the ladder approximation,
that meson-exchange contributions are really needed.
In their replies to the comments on their work,
T\"{o}rnqvist and Roos stressed the concept of {\it duality}.
Hence, it is not even clear whether $t$-channel exchange should be
considered at all, when all propagator modes in $s$-channel exchange
are accounted for.
Duality \cite{JPG2pL167,ARXIV07081983}
probably just works the other way around as well. So
either take all $t$-exchange contributions into account, ideally in an
untruncated fashion and not just in the ladder approximation,
or all $s$-exchange contributions, but not both.

The behavior of the $f_{0}$(980) and $a_{0}$(980) poles,
for variations in the RSE couplings, was studied
in Refs.~\cite{HEPPH0207022,AIPCP688p88}.
No further $t$-channel exchange was needed
in order to describe these scalar resonances
by the lowest dynamically generated poles of the model
in $K\bar{K}$, for respectively $I=0$ and $I=1$.
The modelling is just a bit more complicated
than for the $\sigma$ and the $\kappa$,
because lower lying, but not very strongly coupled,
channels are involved as well,
namely $\pi\pi$ and $\eta\pi$, respectively.
In fact, the correct way to describe the $f_{0}$ resonances
is, of course, by employing a coupled-channel approach to
the $I=0$ $S$-wave
\cite{ZPC30p615,PLB641p265}.
Then one obtains in the RSE both the $f_{0}$(600) ($\sigma$) and
$f_{0}$(980) resonances, in agreement with data, besides other
scattering observables like phase shifts, line-shapes, and inelasticities
\cite{PLB641p265}.

Recently, {\it the true pole positions of broad states}
\/were highlighted
\cite{PRD65p114010,PRL96p132001,EPJC48p553}.
However, pole positions will always depend
on the properties of a specific model.
For narrow resonances there will be agreement,
to some extent, among different models.
But for broad structures it is rather unlikely
that different appoaches will lead to exactly the same pole positions.
Boglione \& Pennington wrote:
{\it ``fitting data along the real axis
cannot accurately determine the true pole position of a broad state
without an analytic continuation, or a very specific model''}
\cite{PRD65p114010}.
Hence, unless we fully agree on the {\it perfect model}
\/to describe the available data,
true pole positions for broad structures do not exist.
At best, we may agree on whether a specific pole exists.

In Ref.~\cite{EPJC22p493},
for the description of elastic $P$-wave $\pi K$ scattering
and the $K^{\ast}(892)$ resonance,
the propagator of Eq.~(\ref{Complitude})
was approximated by the first term
plus a constant representing the remainder of the sum:
\begin{equation}
\sum_{N=0}^{\infty}
\fnd{\abs{{\cal F}_{N}\left( r_{0}\right)}^{2}}
{E-E_{N}}
\longrightarrow\;\fnd{\alpha}{E-E_{0}}-\beta
\;\;\; .
\label{BWKpiPwave}
\end{equation}
Such a procedure is equivalent to the approaches
of T\"{o}rnqvist \cite{ZPC68p647}
and Oset \& Oller \cite{PRD60p074023},
as discussed in the introduction.
In the latter approaches, however, one looses track
of the relations among all coupling constants,
thus needing to introduce arbitrary parameters,
like $\alpha$ and $\beta$ in Eq.~(\ref{BWKpiPwave}),
while also contact with different flavors
and other angular momenta is lost.
Nevertheless, in the approximation of Eq.~(\ref{BWKpiPwave}),
one can study sufficiently well the properties
of the dynamically generated and the lowest $q\bar{q}$
resonances.

In the case of elastic $P$-wave $\pi K$ scattering,
no dynamically generated resonance is found.
So, when the overall coupling is decreased,
the $K^{\ast}(892)$ pole returns to $E_{0}$ \cite{EPJC22p493}.
But for non-exotic elastic $S$-wave $\pi K$ scattering,
the dynamically generated $K^{\ast}_{0}(800)$ resonance appears,
besides the $K^{\ast}_{0}(1430)$.
We may thus conclude that the contact term
indeed absorps those terms of the Regge propagator
which are not accounted for.
In the loop sum~(\ref{Texpansion})
its contribution is negligible,
provided a complete Regge propagator is exchanged.
Hence, this result seems to suggest
that the contact term is not necessary at all in a microscopic
formulation.

The effect of hadron loops on the spectra of mesons and baryons
has been studied by various groups,
and for a variety of different confinement mechanisms
\cite{PRL36p500,AP123p1,AIPCP717p665,PRD50p6855}.
For mesons, the procedure usually amounts to the inclusion of meson loops
in a $q\bar{q}$ description, or, equivalently,
the inclusion of quark loops in a model for meson-meson scattering.
This results in resonance widths,
central masses that do not coincide with the pure confinement spectrum,
mass shifts of bound states,
resonance line-shapes that are very different from the usual Breit-Wigner
ones, threshold effects and cusps.
In particular, it should be mentioned that
mass shifts are large and negative for the ground states
of the various flavor configurations \cite{PRD27p1527}.
Unquenching the lattice is still in its infancy, at least for the light
scalars, as we conclude from Ref.~\cite{HEPLAT0510066}.
However, its effects should not be underestimated.
Hence, ground-state levels of quenched approximations
for $q\bar{q}$ configurations in relative $S$-waves
must be expected to come out far above the experimental masses.

\section{Quark-interchange contributions}

As mentioned in the previous section,
the model is not limited to the study of non-exotic channels.
However, $^{3\!}P_{0}$ pair creation/annihilation,
through which process meson pairs couple to the Regge propagator,
does not work for exotic channels.
Nevertheless, the alternative, which is {\it quark interchange},
does give contributions
to all possible hadronic final-state interactions,
hence also for exotic two-meson channels.

Quark-interchange contributions to meson-meson interactions
can be determined \cite{ZPC38p453} with the very same techniques
that were developed in
Refs.~\cite{PRD25p2406,ZPC17p135,ZPC21p291}.

\section{Experimental results for the Regge spectrum}
\label{production}

As may be concluded from Eq.~(\ref{Texpansion}),
and from the expressions for $V_{\ell}$ and  $\Omega_{\ell}$
given in Eq.~(\ref{Vandomega}),
the dressed partial-wave RSE propagator
for strong interactions takes the form
(restricted to the one-channel case and
leaving out some parts not essential
for our discussion in this section)
\begin{equation}
\bm{\Pi}_{\ell}(E)=\left\{
1-ij_{\ell}\left( pr_{0}\right)
h^{(1)}_{\ell}\left( pr_{0}\right)
\dissum{n=0}{\infty}
\fndrs{5pt}{\abs{g_{NL}}^{2}}{-2pt}{E-E_{NL}}
\right\}^{-1}
\;\;\; .
\label{propagator}
\end{equation}

This propagator has the very intriguing property
that it vanishes for $E\to E_{NL}$.
Hence, one may wonder what happens in a physical process
when the propagator
does not allow any signal to pass.
We shall show in the following that this phenomenon
can be, and has indeed been, observed in experiment,
but not in scattering processes.

The RSE amplitude for strong scattering
is given in Eqs.~(\ref{Texpansion},\ref{uniT}).
One easily verifies that it does not vanish
in the limit $E\to E_{NL}$.
However, for strong production processes,
we deduced in Ref.~\cite{AP323p1215},
following a similar procedure as Roca, Palomar, Oset, and Chiang
in Ref.~\cite{NPA744p127},
a relation between the production amplitude \bm{P}
and the scattering amplitude \bm{T}, reading
\begin{equation}
\bm{P}_{\ell}=
j_{\ell}\left( pr_{0}\right)
+i\,\bm{T}_{\ell}h^{(1)}_{\ell}\left( pr_{0}\right)
\;\;\; ,
\end{equation}
which, using Eqs.~(\ref{propagator}) and (\ref{Texpansion}),
can also be written as
\begin{equation}
\bm{P}_{\ell}=
j_{\ell}\left( pr_{0}\right)
\bm{\Pi}_{\ell}(E)
\;\;\; .
\label{Production}
\end{equation}
For the latter expression we find,
by the use of Eq.~(\ref{propagator}),
that the production amplitude of Eq.~(\ref{Production})
tends to zero when $E\to E_{NL}$.
This effect must be visible in experimental
strong production cross sections.

Actually, the primary question here
is not so much if a vanishing
$q\bar{q}$ propagator is observable,
but rather
whether the production amplitude always vanishes when $E\to E_{NL}$.
In order to answer this,
we must return to the results of Ref.~\cite{AP323p1215},
where we found, for the complete production amplitude
in the case of multi-channel processes,
that Eq.~(\ref{Production}) represents the leading term,
and that the remainder is expressed
in terms of the inelastic components of the scattering amplitude.
The latter terms do not vanish in the limit $E\to E_{NL}$,
as we have discussed above.
Hence, the production amplitude only
vanishes {\it approximately} \/in this limit,
in case inelasticity is suppressed.

However, there are more questions to be responded
with respect to the observability of vanishing propagators.
Namely, do processes exist where only one partial wave contributes
and in which processes other than $s$-channel exchange
do not play an important role?
Fortunately, the answer to the latter, pertinent, question
can be responded afirmatively,
because electron-positron annihilation
into multi-hadron final states
takes basically place via one photon,
hence with $J^{PC}=1^{--}$ quantum numbers.
Consequently, when the photon materializes into
a pair of current quarks,
which couple via the $q\bar{q}$ propagator
to the final multi-hadron state,
we may assume that the intermediate propagator
carries the quantum numbers of the photon.
Moreover, alternative processes are suppressed.

We may thus conclude that,
if we want to discover
whether the propagator really vanishes at $E\to E_{NL}$,
then the ideal touchstone is
$e^{+}e^{-}$ annihilation into multi-hadron states.
But not only do we have at our disposal a wealth of
experimental results on such processes,
there also exist predictions for the values of $E_{NL}$,
with $L=0$ or $L=2$,
given by the parameter set of Ref.~\cite{PRD27p1527}.
As an example, for $c\bar{c}$ one finds
in the latter paper
$E_{0,0}=3.409$ GeV and $\omega=0.19$ GeV.
Using Eq.~(\ref{HOspectrum}), we then get for the higher
$c\bar{c}$ confinement states the spectrum
$E_{1,0}=E_{0,2}=3.789$ GeV,
$E_{2,0}=E_{1,2}=4.169$ GeV,
$E_{3,0}=E_{2,2}=4.549$ GeV, \ldots .

The latter two levels of the $c\bar{c}$ confinement spectrum
can indeed be clearly observed in experiment.
For example, the non-resonant signal in
$e^{+}e^{-}\to\pi^{+}\pi^{-}\psi (2S)$
(see Fig.~5 of Ref.~\cite{PRL99p142002})
is divided into two substructures
\cite{PRD77p014033,PRD78p014032,PLB665p26},
since the full $c\bar{c}$ propagator (\ref{propagator}),
dressed with meson loops,
vanishes at $E_{3}=4.55$ GeV
\cite{PRD27p1527}.
In the same set of data, one may observe a lower-lying zero
at $E_{2}=4.17$ GeV
\cite{PRD27p1527},
more distinctly visible in the data on
$e^{+}e^{-}\to\pi^{+}\pi^{-}J/\psi$
(see Fig.~3 of Ref.~\cite{PRL99p182004}).
The true $c\bar{c}$ resonances
can be found on the slopes
of the above-mentioned non-resonant structures \cite{HEPPH0605317},
unfortunately with little statistical significance,
if any \cite{PRL95p142001}.

In the light-quark sector,
where low statistics does not result in sufficient accuracy,
one needs some imagination
for the identification of vanishing amplitudes
in $e^{+}e^{-}$ annihilation processes
into multi-hadron states.
Thus, from data given in Ref.~\cite{ZPC62p455},
one may infer that certain four-pion amplitudes vanish at, or near,
the predicted \cite{PRD27p1527} 1.097 GeV $n\bar{n}$ ground state,
and near the predicted 1.477 GeV $n\bar{n}$ excited state.
Some further evidence, also for the predicted higher excited states
in $n\bar{n}$ and $s\bar{s}$, is found in data from
Refs.~\cite{PLB73p226,PLB99p257,PLB99p261,ZPC39p13,PLB551p27}.

So we indeed observe minima in production processes,
which confirm vanishing $q\bar{q}$ propagators.
Moreover, the $q\bar{q}$ confinement spectrum predicted
25 years ago in Ref.~\cite{PRD27p1527}
seems to agree well with experimental observations
for vector mesons.
Accordingly, we expect vector-meson $q\bar{q}$ resonances
associated with each of the Regge states:
one ground state, dominantly in a $q\bar{q}$ $S$-wave,
and two resonances for each of the higher excited Regge states, viz.\
one dominantly in an $S$-wave, and the other mostly in a $D$-wave
(see Fig.~\ref{vectors}).

\section{Conclusions}

We have found that the meson spectrum
can be described by $s$-channel exchange of Regge propagators
in non-exotic meson-meson interactions,
and furthermore that the bound-state and resonance spectrum of mesons
is richer than, and different from,
the underlying Regge spectrum \cite{ZPC30p615,AIPCP1030p219}.
For the latter we generally observe
very regular and equidistant level spacings,
instead of quadratic trajectories.
We also have established a link
between $s$-channel exchange of Regge propagators
and unitarized chiral models constructed
for the study of resonances in $S$-wave meson-meson scattering.
Moreover, a method is given to relate the coupling constants
and seed masses for the latter models.
Moreover, we have indicated how the Regge spectrum
of $q\bar{q}$ propagator modes may be observed in production processes.

Finally, the use of Regge propagators for meson physics
is seen \cite{PRD21p772,PRD27p1527}
not to be restricted to light flavors,
but can, without any further effort,
be extended to heavy quarkonia with the same set of parameters,
i.e., the effective quark masses,
the universal frequency $\omega$,
the overall vertex intensity $\lambda$,
and the average string-breaking distance $r_{0}$.

\section*{Acknowledgements}

We wish to thank P.~C.~Magalh\~{a}es for very fruitful discussions,
essential to develop the ideas outlined in this paper.
This work was supported in part by the {\it Funda\c{c}\~{a}o para a
Ci\^{e}ncia e a Tecnologia} \/of the {\it Minist\'{e}rio da Ci\^{e}ncia,
Tecnologia e Ensino Superior} \/of Portugal, under contract
POCI/\-FP/\-81913/\-2007.

\newcommand{\pubprt}[4]{#1 {\bf #2}, #3 (#4)}
\newcommand{\ertbid}[4]{[Erratum-ibid.~#1 {\bf #2}, #3 (#4)]}
\def\AIPCP{AIP Conf.\ Proc.}
\def\AP{Ann.\ Phys.}
\def\APA{Acta Phys.\ Austriaca}
\def\EPJA{Eur.\ Phys.\ J.\ A}
\def\EPJC{Eur.\ Phys.\ J.\ C}
\def\IJTPGTNO{Int.\ J.\ Theor.\ Phys.\ Group Theor.\ Nonlin.\ Opt.}
\def\JPG{J.\ Phys.\ G}
\def\NC{Nuovo Cim.}
\def\NCA{Nuovo Cim.\ A}
\def\NPA{Nucl.\ Phys.\ A}
\def\NPB{Nucl.\ Phys.\ B}
\def\NPPS{Nucl.\ Phys.\ Proc.\ Suppl.}
\def\PLB{Phys.\ Lett.\ B}
\def\PR{Phys.\ Rev.}
\def\PRC{Phys.\ Rev.\ C}
\def\PRD{Phys.\ Rev.\ D}
\def\PRL{Phys.\ Rev.\ Lett.}
\def\ZPC{Z.\ Phys.\ C}

\end{document}